\def\ko{K_0}

\def\rfr#1{eq. (\ref{#1})}

\def\dert#1#2{\frac{{{d}}{#1}}{{{d}}{#2}}}

\def\virg#1{``#1''}

\def\eqi{\begin{equation}}
\def\eqf{\end{equation}}
\def\eqia{\begin{eqnarray}}
\def\eqfa{\end{eqnarray}}
\def\Om{\mathit{\Omega}}
\def\rp#1#2{{#1\over#2}}
\def\lb#1{\label{#1}}

\def\bds#1{\boldsymbol{#1}}

\def\ee{e^2}

\def\ton#1{\left(#1\right)}
\def\qua#1{\left[#1\right]}

\documentclass[Galaxies,article,accept,oneauthor,dvipdfm,12pt,a4paper]{mdpi} 

\lastpage{x}
\doinum{10.3390/------}
\pubvolume{xx}
\pubyear{2013}
\history{Received: xx / Accepted: xx / Published: xx}
\usepackage{amsmath,amssymb,amsthm,amscd,latexsym}
\usepackage{hyperref}
\usepackage[toc,title,titletoc]{appendix}
\usepackage{graphicx,epsfig}
\RequirePackage{color}

\Title{A Closer Earth and the Faint Young Sun Paradox: Modification of the Laws of Gravitation, or Sun/Earth Mass Losses?}

\Author{Lorenzo Iorio $^{1,}$*}

\address{%
$^{1}$ Italian Ministry of Education, University and Research (M.I.U.R.)-Education, Fellow of the Royal Astronomical Society (F.R.A.S.), Viale Unit\`{a} di Italia 68, 70125, Bari (BA), Italy. Tel. +39 3292399167}

\corres{lorenzo.iorio@libero.it}

\abstract{
Given a solar luminosity $L_{\rm Ar}=0.75 L_0$ at the beginning of the Archean $3.8$ \textcolor{black}{Ga} ago, where $L_0$ is the present-day one, if the heliocentric distance $r$ of the Earth was $r_{\rm Ar}=0.956 r_0$, the solar irradiance would have been as large as $I_{\rm Ar}=0.82 I_0$. It would allowed for a liquid ocean on the terrestrial surface which, otherwise, would have been frozen, contrary to the empirical evidence. By further assuming that some physical mechanism subsequently displaced the Earth towards its current distance in such a way that the irradiance stayed substantially constant  over the entire Archean from $3.8$ \textcolor{black}{Ga} to $2.5$ \textcolor{black}{Ga} ago, a relative recession \textcolor{black}{per year} as large as $\dot r/r \textcolor{black}{ \ \approx\  } 3.4\times 10^{-11}\ {\rm \textcolor{black}{a}^{-1}}$ would have been required. Although such a figure is roughly of the same order of magnitude of the value of the Hubble parameter $3.8$ \textcolor{black}{Ga} ago $H_{\rm Ar}=1.192 H_0=8.2\times 10^{-11}\ {\rm \textcolor{black}{a}^{-1}}$, standard general relativity rules out cosmological explanations for the hypothesized Earth's recession rate. Instead, a class of modified theories of gravitation with nonminimal coupling between the matter and the metric naturally predicts a secular variation of the relative distance of a localized two-body system, thus yielding a potentially viable candidate to explain the putative recession of the Earth\textcolor{black}{'s} orbit. Another competing mechanism of classical origin which could, in principle, allow for the desired effect is the mass loss which either the Sun or the Earth itself may have experienced during the Archean. On the one hand, this implies that our planet should have lost $2\textcolor{black}{\ \%}$ of its present mass in the form of eroded/evaporated hydrosphere.
On the other hand, it is widely believed that the Sun could have lost mass at an enhanced rate due to a stronger solar wind in the past for not more than \textcolor{black}{$ \ \approx\   (0.2\ {\rm to}\ 0.3)$ Ga}.
}

\keyword{Archean period; Paleoclimatology; Solar physics; Experimental studies of gravity; Relativity and gravitation; Modified theories of gravity; Celestial mechanics}

\PACS{91.70.hf; 92.60.Iv; 96.60.-j; 04.80.-y;  95.30.Sf; 04.50.Kd; 95.10.Ce}

\begin{document}

%
\section{Introduction}\lb{intro}

The so-called \virg{Faint Young Sun Paradox} (FYSP)  \citep{1972Sci...177...52S} consists in the fact that, according to consolidated models of the Sun's evolution history, the energy output of our star during the Archean, from $3.8$ \textcolor{black}{Ga} to $2.5$ \textcolor{black}{Ga} ago, would have been too low to keep liquid water on the Earth's surface. \textcolor{black}{Instead,} there are compelling and independent evidences that, actually, our planet was mostly covered by liquid water oceans, hosting also forms of life, during that eon. For a recent review of the FYSP, see \citep{Feul012} and references therein.

The bolometric solar luminosity \textcolor{black}{$L$} measures the electromagnetic radiant power emitted by the Sun integrated over all the wavelengths.
The solar irradiance $I$ measured at the Earth's atmosphere is defined as the ratio of the solar luminosity to the area of a sphere centered on the Sun with  radius equal to the Earth-Sun distance $r$; in the following, we will assume a circular orbit for the Earth. Thus, its current value is \textcolor{black}{\citep{2011GeoRL..38.1706K}}
\eqi I_0= \textcolor{black}{(}1360.8\pm 0.5\textcolor{black}{)}\ {\rm W\ m^{-2}}.\eqf

Setting the origin of the time $t$ at the Zero-Age Main Sequence (ZAMS) epoch, i.e. when the nuclear fusion ignited in the core of the Sun, a formula which accounts for the temporal evolution of the solar luminosity reasonably well over the eons, with the possible exception of the first $\textcolor{black}{ \ \approx\  } 0.2$ \textcolor{black}{Ga} in the life of the young Sun, is \citep{1981SoPh...74...21G}
\eqi \rp{L(t)}{L_0} = \rp{1}{1+\rp{2}{5}\ton{1-\rp{t}{t_0}} },\lb{bolo}\eqf
where $t_0 = 4.57\ {\rm Ga}$ is the present epoch, and $L_0$ is the current Sun's luminosity. See, e.g., Figure 1 in \citep{Feul012}, and Figure 1 in \citep{2010IAUS..264....3R}.
The formula of \rfr{bolo} is in good agreement with recent standard solar models such as, e.g., \citep{2001ApJ...555..990B}.

According to \rfr{bolo}, at the beginning of the Archean era $3.8$ \textcolor{black}{Ga} ago, corresponding to $t_{\rm Ar}=0.77$ \textcolor{black}{Ga} in our ZAMS-based temporal scale, the solar luminosity was
just \eqi L_{\rm Ar} = 0.75 L_0.\lb{lumi}\eqf
Thus, if the heliocentric distance of the Earth was the same as today, \rfr{lumi} implies that
\eqi I_{\rm Ar} = 0.75 I_0.\lb{alura}\eqf
As extensively reviewed in \citep{Feul012}, there is ample and compelling evidence that the Earth hosted  liquid water, and even life, during the entire Archean eon spanning about $1.3$ Ga. Thus, our planet could not be entirely frozen during such a remote eon, as, instead, it would have necessarily been if it really received only $\textcolor{black}{ \ \approx\  } 75\textcolor{black}{\ \%}$ of the current solar irradiance, as it results from \rfr{alura}.

Although intense efforts by several researchers in the last decades to find a satisfactory solution to the FYSP involving multidisciplinary investigations on deep-time paleoclimatology \citep{Driese011}, greenhouse effect \citep{2013Sci...339...64W}, ancient cosmic ray flux \citep{2012ApJ...760...85C}, solar activity \citep{2007LRSP....4....3G} and solar wind \citep{lrsp-2004-2}, it not only  refuses to go away \citep{2010Natur.464..687K,2011Natur.474E...1G,Feul012}, but rather it becomes even more severe \citep{2012GeoRL..3923710K} in view of some recent studies.
\textcolor{black}{This is not to claim that the climatic solutions are nowadays ruled out \citep{2010Natur.464..744R,2013Sci...339...44K}, especially those involving a carbon-dioxide greenhouse in the early Archean and a carbon dioxide-methane greenhouse at later times \citep{2008AsBio...8.1127H,2013Sci...339...64W,Schultz013}; simply, we feel that it is worthwhile pursuing also different lines of research.}

In this paper, as a preliminary working hypothesis, we consider the possibility that the early Earth was closer to the Sun just enough to keep liquid oceans on its surface during the entire Archean eon. For other, more or less detailed, investigations  in the literature along this line of research, see \citep{2000GeoRL..27..501G,2007ApJ...660.1700M,Feul012,pazzo}. In Section \ref{vicina}, we explore the consequences of such an assumption from a phenomenological point of view.  After critically reviewing in Section \ref{cosmic} some unsatisfactorily attempts of cosmological origin to find an explanation for the required orbital recession of our planet, we offer some hints towards a possible solution both in terms of fundamental physics (Section \ref{fundam}) and by considering certain partially neglected classical orbital effects due to possible mass loss rates potentially experienced by the Sun and/or the Earth in the Archean (Section \ref{massloss}). Section \ref{conclude} summarize\textcolor{black}{s} our findings.
\section{A working hypothesis: was the Earth closer to the Sun than now?}\lb{vicina}

As a working hypothesis, let us provisionally assume that, at $t_{\rm Ar}$, the solar irradiance $I_{\rm Ar}$  was approximately equal to a fraction  of the present one $I_0$ large enough to allow for a global liquid ocean on the Earth. As noticed in \citep{Feul012},
\textcolor{black}{earlier studies \citep{1969Tell...21..611B,1969JApMe...8..392S,1992GPC.....5..133G}} required an Archean luminosity as large as \textcolor{black}{$98\ \%$ to $85\ \%$} of the present-day value to have liquid water.
\textcolor{black}{Some more recent} models have lowered the critical luminosity threshold  down to about \textcolor{black}{$90\ \%$ to $86\ \%$} \citep{1993JGR....9820803J,1997GPC....14...97L}, with a lower limit as little as \citep{1997GPC....14...97L} \eqi L_{\rm oc}\textcolor{black}{ \ \approx\  } 0.82 L_0.\lb{oce}\eqf   Since  the same heliocentric distance as the present-day one was assumed in the literature, \rfr{oce} is equivalent to the following condition for the irradiance required to keep liquid ocean \eqi I_{\rm oc} \textcolor{black}{ \ \approx\  } 0.82 I_0\lb{irra}.\eqf By assuming $I_{\rm Ar} = I_{\rm oc}$, together with \rfr{lumi},
implies
\eqi r_{\rm Ar} = 0.956 r_0,\lb{distanza}\eqf i.e. the Earth should have been closer to the Sun by about $\textcolor{black}{ \ \approx\  } 4.4\textcolor{black}{\ \%}$ with respect to the present epoch.
As a consequence, \textcolor{black}{if one assumes that the FYSP could only be resolved by a closer Earth}, some physical mechanism \textcolor{black}{should} have subsequently displaced out planet to roughly its current heliocentric  distance by keeping the irradiance equal to at least $I_{\rm oc}$ over the next $1.3$ \textcolor{black}{Ga} until the beginning of the Proterozoic era $2.5$ \textcolor{black}{Ga} ago, corresponding to
$t_{\rm Pr} = 2.07$ \textcolor{black}{Ga} with respect to the ZAMS epoch, when the luminosity of the Sun was \eqi L_{\rm Pr}= 0.82 L_0,\eqf according to \rfr{bolo}.
Thus, by imposing
\eqi I(t) = 0.82 I_0,\  0.77\ {\rm Ga} \leq t\leq 2.07\ {\rm Ga},\eqf
one gets
\begin{align}
\rp{r(t)}{r_0} \lb{erre}& = \rp{1}{\sqrt{0.82\qua{1 + \rp{2}{5}\ton{1-\rp{t}{t_0}}  }}}, \\ \nonumber \\
\rp{\dot r(t)}{r(t)} \lb{erredot}& = \rp{1}{7t_0\ton{1 - \rp{2}{7}\rp{t}{t_0}}}.
\end{align}
The plots of \rfr{erre}-\rfr{erredot} are depicted in Figure \ref{figura1}.
\begin{figure*}
\centering
\begin{tabular}{c}
\epsfig{file=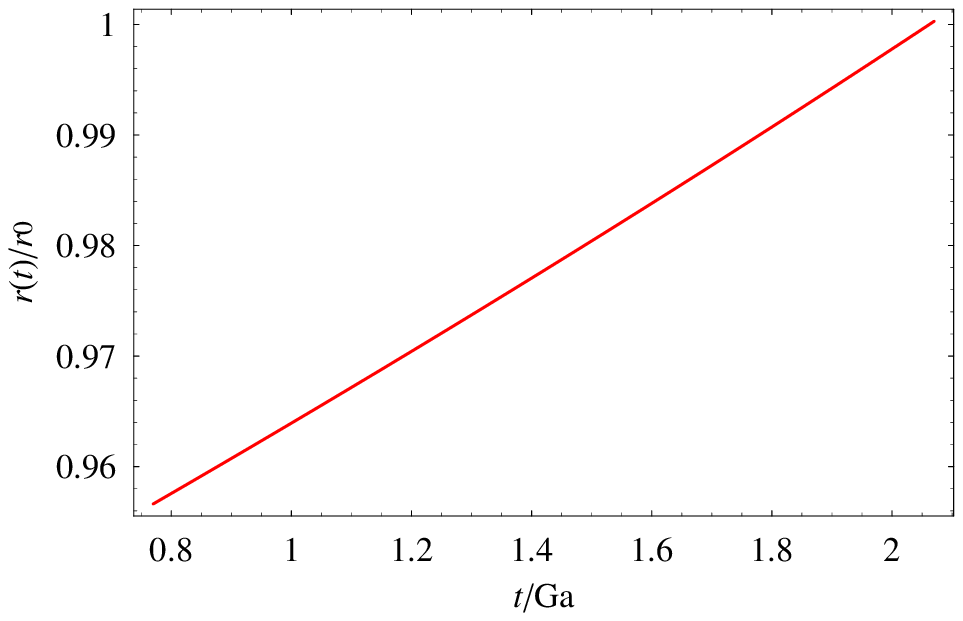,width=0.50\linewidth,clip=}\\
\epsfig{file=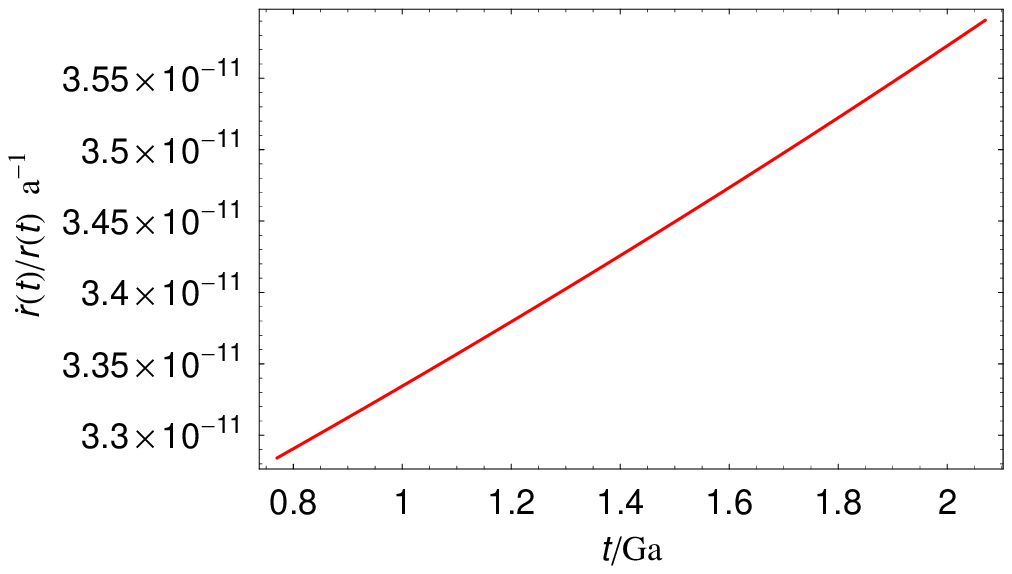,width=0.50\linewidth,clip=}\\
\end{tabular}
\caption{
Upper panel: temporal evolution of the Earth-Sun distance $r(t)$, normalized to its present-day value $r_0$, over the Archean according to \rfr{erre}. Lower panel: temporal evolution of $\dot r(t)/r(t)$ over the Archean according to \rfr{erredot}. In both cases the constraint $I(t)=0.82 I_0$ throughout the Archean was adopted.
}\lb{figura1}
\end{figure*}
It can be noticed that a percent distance rate as large as
\eqi\rp{\dot r}{r}\textcolor{black}{ \ \approx\  } 3.4\times 10^{-11}\ {\rm \textcolor{black}{a}}^{-1}\lb{rate}\eqf
is enough to keep the irradiance equal to about $82\textcolor{black}{\ \%}$ of the present one during the entire Archean by displacing the Earth towards its current location.

A very important point is to search for independent evidences supporting or contradicting the hypothesis of a closer Earth at the beginning of the Archean. From the third Kepler law of classical gravitational physics, it turns out that, if the heliocentric distance of the Earth was smaller, then the duration of the year should have been \textcolor{black}{shorter}. As remarked in \citep{2007ApJ...660.1700M}, in principle, a \textcolor{black}{shorter} terrestrial year should have left traces in certain geological records such as tidal rhythmites and banded iron formations. Actually, the available precision of such potentially interesting indicators is rather poor for pre-Cambrian epochs \citep{2000RvGeo..38...37W,2000Geo....28..831E} to draw any meaningful conclusion. However, it cannot rule out our hypothesis.

Finally, we wish to mention that a hypothesis somewhat analogous to that presented here was proposed in \citep{2010ChSBu..55.4010Z}, although within a different temporal context. Indeed,  in \citep{2010ChSBu..55.4010Z}, by analyzing the measurements of the growth patterns on fossil corals, it was claimed that, at the beginning of the Phanerozoic eon $0.53$ \textcolor{black}{Ga} ago, it was $r_{\rm Ph}=0.976r_0$
\section{Ruling out cosmological explanations}\lb{cosmic}
\subsection{\textcolor{black}{The accelerated expansion of the Universe}}

Given the timescales involved in such processes, it is worthwhile investigating if such a putative recessions of the Earth's orbit could be induced by some effects of cosmological nature, as preliminarily suggested in \citep{2010ChSBu..55.4010Z,2012NewA...17....1K}. Such a possibility is made appealing by noticing that the rate in \rfr{rate} is of the same order of magnitude of  the currently accepted value of the Hubble parameter \citep{2013arXiv1303.5076P}
\eqi H_0 = \textcolor{black}{(}67.4\pm 1.4\textcolor{black}{)}\ {\rm km\ s^{-1}\ Mpc^{-1}} = (6.89\pm 0.14 )\times 10^{-11}\ {\rm \textcolor{black}{a}^{-1}}.\lb{bubble}\eqf
The Hubble parameter is defined as \citep{Pad010}
\eqi H\ton{t}\doteq \rp{\dot S(t)}{S(t)},\lb{hu}\eqf where $S(t)$ is the cosmological expansion factor; the definition of \rfr{hu} is valid at any time $t$.
As a  first step of our inquiry, an accurate calculation of the value of the Hubble parameter $3.8$ \textcolor{black}{Ga} ago, accounting for the currently accepted knowledge of the cosmic evolution, is required.
Let us briefly recall that  the simplest cosmological model providing a reasonably good match to several different types of  observations is the so-called
$\Lambda$CDM model; in addition to the standard forms of baryonic matter and radiation, it also implies the existence of the dark energy, accounted for by a cosmological constant $\Lambda$, and of the non-baryonic cold dark matter. It relies upon the general relativity by Einstein as the correct theory of the gravitational interaction at cosmological scales.
The first Friedmann equation for a Friedmann-Lema\^{\i}tre-Roberston-Walker (FLRW) spacetime metric describing a homogenous and isotropic non-empty Universe endowed with a cosmological constant $\Lambda$ is \citep{Pad010}
\eqi \ton{\rp{\dot S}{S}}^2 + \rp{k}{S^2}= H_0^2\qua{\Omega_{\rm R}\ton{\rp{S_0}{S}}^4 + \Omega_{\rm NR}\ton{\rp{S_0}{S}}^3+\Omega_{\Lambda}},\lb{lunga}\eqf
where $k$ characterizes the curvature of the spatial hypersurfaces, $S_0$ is the present-day value of  the expansion factor, and the dimensionless energy densities $\Omega_{i}, i={\rm R, NR},\Lambda$, normalized to the critical energy density \eqi\varepsilon_{\rm c} = \rp{3c^2 H_0^2}{8\pi G},\eqf where $G$ is the Newtonian constant of gravitation and $c$ is the speed of light in vacuum,  refer to their values at $S = S_0$. Based on the equation of state relating the pressure $p$ to the energy density $\varepsilon$ of each component, $\Omega_{\rm R}$ refers to the relativistic matter characterized by $p_{\rm R}=(1/3)\varepsilon_{\rm R}$, $\Omega_{\rm NR}$ is the sum of  the normalized energy densities of the ordinary baryonic matter and of the non-baryonic dark matter, both non-relativistic, while $\Omega_{\Lambda}$ accounts for the dark energy modeled by the cosmological constant $\Lambda$ in such a way that $p_{\Lambda} = -\varepsilon_{\Lambda}$.
By keeping only $\Omega_{\rm NR}$ and $\Omega_{\Lambda}$ in \rfr{lunga}, it is possible to integrate it, with $k=0$, to determine $S(t)$. The result is \citep{Pad010}
\eqi \rp{S(t)}{S_0} = \ton{\rp{\Omega_{\rm NR}}{\Omega_{\Lambda}}}^{1/3}\sinh^{2/3}\ton{\rp{3}{2}\sqrt{\Omega_{\Lambda}}H_0 t}, \ \Omega_{\rm NR}+\Omega_{\Lambda}=1\lb{Esse}.\eqf
For the beginning of the Archean eon, $3.8$ \textcolor{black}{Ga} ago,  \rfr{Esse} yields
\eqi \rp{S_{\rm Ar}}{S_0} =0.753.\lb{kizza}\eqf Note that in \rfr{Esse} $t$ is meant to be counted from the Big-Bang singularity in such a way that the present epoch is \citep{2013arXiv1303.5076P}
\eqi  t_0  = \textcolor{black}{(}13.813\pm 0.058\textcolor{black}{)}\ {\rm Ga};\lb{epoca}\eqf thus, \eqi t_{\rm Ar}=10.013\ {\rm Ga}\lb{tAR}\eqf has to be used in \rfr{Esse} to yield \rfr{kizza}. Incidentally, from \rfr{bubble} and \rfr{epoca} it turns out
\eqi H_0 t_0 = 0.952.\eqf
As a consequence of \rfr{kizza}, the dimensionless redshift parameter at the beginning of the Archean is
\eqi z_{\rm Ar} =\rp{S_0}{S_{\rm Ar}} - 1=0.32.\eqf It is, thus, a-posteriori confirmed the validity of using \rfr{Esse} for our purposes  since, according to Type Ia supernov{\ae} (SNe Ia) data analyses, the cosmic acceleration started at \citep{2004ApJ...607..665R} \eqi z_{\rm acc} = 0.43\pm 0.13,\eqf
corresponding to about $5.6$ \textcolor{black}{Ga} to $3.5$ \textcolor{black}{Ga} ago.
From \rfr{hu} and \rfr{Esse}, it can be obtained
\eqi H(t) = H_0\sqrt{\Omega_{\Lambda}}\coth\ton{\rp{3}{2}\sqrt{\Omega_{\Lambda}}H_0 t}.\lb{Acca}\eqf
Since \citep{2013arXiv1303.5076P}
\begin{align}\Omega_{\Lambda}&=0.686\pm 0.020,\\ \nonumber \\
 %
 %
 \end{align}
  \rfr{Acca}, together with \rfr{bubble} and \rfr{tAR}, yields
 \eqi H_{\rm Ar} = 1.192 H_0 \lb{Har}\eqf for the value of the Hubble parameter at the beginning of the Archean eon.
At this point, it must be noticed that \rfr{rate} differs from \rfr{Har} at a $\textcolor{black}{ \ \approx\  } 30\sigma$ level.

Even putting aside such a numerical argument, there are also sound theoretical reasons to discard a cosmological origin for the putative secular increase of the Sun-Earth distance at some epoch such as, e.g., the Archean or the Phanerozoic.
It must be stressed that having at disposal the analytical expression of the test particle acceleration caused by a modification of the standard two-body laws of motion more or less deeply rooted in some cosmological scenarios is generally not enough. Indeed, it must explicitly be shown that such a putative cosmological acceleration is actually capable to induce a secular variation of the distance of the test particle with respect to the primary. In fact, it is not the case just for some potentially relevant accelerations of cosmological origin which, instead, have an impact on different features of the two-body orbital motion such as, e.g., the pericenter $\omega$, etc. Standard general relativity predicts that, at the Newtonian level, no two-body acceleration of the order of $H$ occurs
\citep{2010RvMP...82..169C, 2013arXiv1306.0374G}. At the Newtonian level, the first non-vanishing effects of the  cosmic expansion are of the order of $H^2$ \citep{2010RvMP...82..169C, 2013arXiv1306.0374G}; nonetheless, they do not secularly affect the mean distance  of a test particle with respect to the primary since they are caused by an additional radial acceleration proportional to the two-body position vector $\bds r$ which only induces a secular precession of the pericenter  of the orbit \citep{2007PhRvD..75h2001A}. At the post-Newtonian level, a cosmological acceleration of the order of $H$ and proportional to the orbital velocity $\bds v$ of the test particle has recently been found \citep{2012PhRvD..86f4004K}. In principle, it is potentially interesting since it secularly affects both the semimajor axis $a$ and the eccentricity $e$ in such a way that $\overline{r}=a\ton{1+\ee/2}$ changes as well \citep{2013MNRAS.429..915I}. Nonetheless, its percent rate of change  is far too small since it is proportional to $H\ton{v/c}^2$ \citep{2013MNRAS.429..915I}.
As far as the acceleration of the cosmic expansion, driven by the cosmological constant $\Lambda$, is concerned, it only affects the local dynamics of a test particle through a pericenter precession, leaving both $a$ and $e$ unaffected \citep{2003CQGra..20.2727K}.
\subsection{\textcolor{black}{A time-dependent varying gravitational parameter $G$}}

\textcolor{black}{The possibility that the Newtonian coupling parameter $G$ may decrease in time in accordance with  the expansion of the Universe dates back to the pioneeristic studies by Milne \citep{1935QB500.M5.......,1937RSPSA.158..324M}, Dirac \citep{1937Natur.139..323D},  Jordan \citep{1937NW.....25..513J}.
As a consequence, also the dynamics of a two-body system would be affected according to
\eqi\rp{\dot r\ton{t}}{r\ton{t}} = -\rp{\dot G\ton{t}}{G\ton{t}}.\eqf
}

\textcolor{black}{Nonetheless, the present-day bounds on the percent variation rate of $G$ \citep{2004PhRvL..93z1101W,Mueller07}
\begin{align}
\left|\rp{\dot G}{G}\right|  \lb{vincoli1} & \leq 7\times 10^{-13}\ {\rm \textcolor{black}{a}^{-1}}, \\ \nonumber \\
\left|\rp{\dot G}{G}\right| \lb{vincoli2} & \leq 9\times 10^{-13}\ {\rm \textcolor{black}{a}^{-1}},
\end{align}
inferred from the analysis of multidecadal records of observations performed with the accurate Lunar Laser Ranging (LLR) technique \citep{Murphy013}, are  smaller than \rfr{rate} by two orders of magnitude. It could be argued that, after all, the constraints of \rfr{vincoli1}-\rfr{vincoli2} were obtained from data covering just relatively few years if compared with the timescales we are interested in. Actually, in view of the fundamental role played by $G$, its putative variations would have a decisive impact on quite different phenomena such as the evolution of the Sun itself, ages of globular clusters, solar and stellar seismology, the Cosmic Microwave Background (CMB), the Big Bang Nucleosynthesis (BBN), etc.; for a comprehensive review, see \citep{2011LRR....14....2U}. From them, independent constraints on $\dot G/G$, spanning extremely wide timescales, can be inferred. As it results from Sect. 4 of \citep{2011LRR....14....2U}, most of the deep-time ones are $2-3$ orders of magnitude smaller than \rfr{rate}.
}
\section{Unconventional orbital effects}\lb{fundam}

\subsection{Modified gravitational theories with nonminimal coupling}

If standard general relativity does not predict notable cosmological effects able to expand the orbit of a localized two-body system,  it can be done by a certain class \citep{2013PhRvD..87d4045P} of modified gravitational theories with nonminimal coupling between the matter and the gravitational field \citep{1984FoPh...14..865G}. This is not the place to delve into the technical details of such alternative theories of gravitation predicting a violation of the equivalence principle \citep{1984FoPh...14..865G,2011PhRvD..84f4022B,2013PhRvD..87d4045P}. Suffice it to say that a class of them, recently investigated in \citep{2013PhRvD..87d4045P},  yields an extra-acceleration ${\bds A}_{\rm nmc}$ for a test particle orbiting a central body which, interestingly, has a long-term impact on its distance.

In the usual four-dimensional spacetime language, a non-geodesic four-acceleration of a non-rotating test particle  \citep{2013PhRvD..87d4045P}
\eqi A^{\mu}_{\rm nmc} = \rp{c\xi}{\mathfrak{m}}\ton{\delta^{\mu}_{\nu} -\rp{v^{\mu}v_{\nu}}{c^2}}K^{\nu},\ \mu=0,1,2,3\lb{4accel}\eqf occurs.
We adopt the convention according to which the Greek letters are for the spacetime indices, while the Latin letters denotes the three-dimensional spatial indices; in \citep{2013PhRvD..87d4045P} the opposite convention is followed.
In \rfr{4accel}, $\mathfrak{m}$ is the mass of the test particle as defined in multipolar schemes in the context of general relativity,  $\delta_{\nu}^{\mu}$ is the Kronecker delta in four dimensions, $v^{\mu}, v_{\nu}$ are the contravariant and covariant components of the the four-velocity of the test particle, respectively, ($v_{0} = v^0, v_i = -v^i, i=1,2,3$), $\xi$ is an integrated quantity depending on the matter distribution of the system, $K^{\mu}\doteq \nabla^{\mu}\ln F,$ where $\nabla^{\mu}$ denotes the covariant derivative, and the nonminimal function $F$ depends arbitrarily on the spacetime metric $g_{\mu\nu}$ and on the Riemann curvature tensor $R_{\mu\nu\alpha}^{\ \ \ \ \beta}$.
From \rfr{4accel}, the test particle acceleration
\eqi\bds{A}_{\rm nmc} =  -\rp{\xi\qua{c^2 \bds K - c\ko \bds v +\ton{\bds K\bds\cdot\bds v}\bds v }}{c \mathfrak{m}}\lb{accel},\eqf written in the usual three-vector notation, can be extracted.
%
%
%
%
%
In deriving \rfr{accel}, we assumed the slow-motion approximation  in such a way that $v^\mu \textcolor{black}{ \ \approx\  } \{c,\bds v\}$.

A straightforward but cumbersome perturbative calculation can be performed with the standard Gauss equations for the variation of the Keplerian orbital elements \citep{Bertotti03}, implying the decomposition of \rfr{accel} along the radial, transverse and normal directions of an orthonormal trihedron comoving with the particle and their evaluation onto a Keplerian ellipse, usually adopted as unperturbed reference trajectory. Such a procedure, which has the advantage of being applicable to whatsoever perturbing acceleration, yields, to zero order in the eccentricity $e$ of the test particle, the following percent secular variation of its semimajor axis
\eqi\rp{\dot a}{a} = \rp{2\xi K_0}{\mathfrak{m}} + \mathcal{O}\ton{e}.\lb{dadt} \eqf It must be stressed that, for the quite general class of theories covered in \citep{2013PhRvD..87d4045P}, $\mathfrak{m},\xi,K_0$ are, in general, not constant. As a working hypothesis, in obtaining \rfr{dadt} we assumed that they can be considered constant over the  period of the test particle. Thus, there is still room for a slow temporal dependence with characteristic time scales quite larger that the test particle's period. Such a feature  is important to explain the fact that, at present, there is no evidence for any anomalous increase of the Sun-Earth distance as large as a few meters per year, as it would be required by \rfr{rate}. Indeed, it can always be  postulated that, in the last $\textcolor{black}{ \ \approx\  } 2$ Ga, $\mathfrak{m},\xi,K_0$ became smaller enough to yield effects below the current threshold of detectability which, on the basis of the results in \citep{pitjeva07},  was evaluated to be of the order of \citep{2013MNRAS.429..915I} $\textcolor{black}{ \ \approx\  } 1.5\times 10^{-2}\ {\rm m\ \textcolor{black}{a}^{-1}}$ for the Earth. The rate of change of \rfr{dadt} is an important result since it yields an effect  which is rooted in a well defined theoretical framework. It also envisages the exciting possibility that a modification of the currently accepted laws of the gravitational interaction can, in principle, have an impact on the ancient history of our planet and, indirectly, even on the evolution of the life on it.

\subsection{The secular increase of the astronomical unit}

At this point, the reader may wonder why, in the context of a putative increase of the radius
of the Earth's orbit, no reference has been made so far \textcolor{black}{to its secular increase reported by} \cite{2004CeMDA..90..267K,2005tvnv.conf..163S,2009IAU...261.0702A} whose \textcolor{black}{rate} ranges from $ \ \approx\   1.5\times 10^{-1}$ m \textcolor{black}{a}$^{-1}$ \cite{2004CeMDA..90..267K} to $ \ \approx\   5\times 10^{-2}$ m \textcolor{black}{a}$^{-1}$ \cite{2005tvnv.conf..163S}.  Actually, if steadily projected backward in time until $t_{\rm Ar}$, the figures for its secular rate present in the literature would yield a displacement of the Earth's orbit over the last $3.8$ \textcolor{black}{Ga} as little as $\Delta r \ \approx\   (2\ {\rm to}\ 6)\times 10^8\ {\rm m}$, corresponding to $\textcolor{black}{ \ \approx\   (1\ {\rm to}\ 4)\times 10^{-3}  r_0}$, contrary to \rfr{distanza}.
%
\section{Some non-climatic, classical orbital effects}\lb{massloss}

It is important to point out that, actually, there are also some standard physical phenomena which, in principle, could yield a cumulative widening of the Earth's orbit.

\subsection{Gravitational billiard}

It was recently proposed \citep{pazzo} that our planet would have migrating to its current distance in the Archean as a consequence of a gravitational billiard involving  planet-planet scattering between the Earth itself and a rogue rocky protoplanetesimal X, with $m_{\rm X} \textcolor{black}{ \ \approx\  } 0.75 m_{\oplus}$, which would have impacted on Venus. However, as the author himself of \citep{pazzo} acknowledges, \virg{this may not be compelling in the face of minimal constraints}.

\subsection{Mass losses}

Another classical effect, for which independent confirmations in several astronomical scenarios exist, is the mass loss of main sequence stars \citep{2007A&A...463...11H} and/or of the surrounding planets due to the possible erosion of their hydrospheres/atmospheres \citep{2011A&A...529A.136E} caused by the stellar winds \citep{lrsp-2004-2,SuzukiEPS2012}. Their gravitational effects on the dynamics of a two-body system have been worked out in a number of papers in the literature, especially as far as the mass loss of the hosting star is concerned; see, e.g., \citep{2010NatSc...2..329I,2013MNRAS.432..438A,2013MNRAS.431.2971L} and references therein. In regard to the orbital recession of a planet losing mass because of  the stellar wind of its parent star, see \citep{2012NewA...17..356I} and references therein. Let us explore the possibility that, either partly or entirely, they can account for the phenomenology described in Section \ref{vicina} within our working hypothesis of a closer Earth $3.8$ \textcolor{black}{Ga} ago. For a previous analysis involving only the Sun's mass loss, see \citep{2007ApJ...660.1700M}.
\subsubsection{Isotropic mass loss of the Sun}

As far as the Sun is concerned, it is believed that, due to its stronger activity in the past \citep{lrsp-2004-2,Feul012} associated with faster rotation and stronger magnetic fields, its mass loss rate driven by the solar wind was higher \citep{SuzukiEPS2012} than the present-day one \citep{2013MNRAS.tmp.1343P}
\eqi \left.\rp{\dot M_{\odot}}{M_{\odot}}\right|_0 = (-6.3\pm 4.3)\times 10^{-14}\ {\rm \textcolor{black}{a}^{-1}},\lb{ora}\eqf
recently measured in a model-independent way from the planetary orbital dynamics.
Since from the cited literature it turns out that
\eqi\rp{\dot r(t)}{r(t)} = -\rp{\dot M_{\odot}(t)}{M_{\odot}(t)},\lb{ipotesi}\eqf
\rfr{rate} tells us that a steady solar mass loss rate as large as \eqi\rp{\dot M_{\odot}}{M_{\odot}}\textcolor{black}{ \ \approx\   (3.4\ {\rm to}\ 3.5)}\times 10^{-11}\ {\rm \textcolor{black}{a}}^{-1},\ 0.77\ {\rm Ga}\leq t\leq 2.07\ {\rm Ga} \lb{necess}\eqf
would be needed if it was to be considered as the sole cause for the  increase of the size of the Earth's orbit hypothesized in \rfr{erredot}.
It is interesting to compare our quantitative estimate in \rfr{necess} with the order-of-magnitude estimate in  \citep{2000GeoRL..27..501G} pointing towards a mass loss rate of the order of \eqi\rp{\dot M_{\odot}}{M_{\odot}}\textcolor{black}{ \ \approx\   (10^{-11}\ {\rm to}\ 10^{-10})}\ {\rm \textcolor{black}{a}}^{-1}.\eqf
See also \citep{2007ApJ...660.1700M}.
It is worthwhile noticing that  \rfr{necess} implies \eqi M_{\odot}^{\rm Ar}\textcolor{black}{ \ \approx\  } 1.044 M^0_{\odot}.\lb{massa}\eqf In principle, \rfr{massa} may contradict some of the assumptions on which the reasoning of Section \ref{vicina}, yielding  just \rfr{rate} and Figure \ref{figura1}, is based. Indeed, the luminosity of a star powered by nuclear fusion is proportional to \citep{1994sse..book.....K}
\eqi L\propto M^{\eta}, \ 2\lesssim \eta \lesssim 4,\eqf with $\eta=\eta(M)$; for a Sun-like star, it is $\eta \textcolor{black}{ \ \approx\  } 4$. Thus, by keeping \rfr{distanza} for $r_{\rm Ar}$, it would be
\begin{align}
L_{\rm Ar}&\textcolor{black}{ \ \approx\  } 0.84 L_0,\\ \nonumber \\
I_{\rm Ar} \lb{contraz}&\textcolor{black}{ \ \approx\  } 0.92 I_0.
\end{align}
However, it may be that the uncertainties in \rfr{bolo} and, especially, in $\eta$ might reduce the discrepancy between \rfr{irra} and \rfr{contraz}.
On the other hand, we also mention the fact that a Sun's mass larger by just $4.4\textcolor{black}{\ \%}$   would not pose the problems mentioned in Section 4 of \citep{Feul012} concerning the evaporation of the terrestrial hydrosphere.
In fact, the actual possibility that the Sun may have experienced a reduction of its mass such as the one postulated in \rfr{necess} should be regarded as somewhat controversial, as far as both the timescale and the magnitude itself of the solar mass loss rate are concerned \citep{2007ApJ...660.1700M}. Indeed, Figure 15 of \citep{lrsp-2004-2} indicates a Sun's mass loss rate smaller than \rfr{ipotesi} by about \textcolor{black}{one to two} orders of magnitude during the Archean, with a maximum of roughly \eqi\left.\rp{\dot M_{\odot}}{M_{\odot}}\right|_{\rm Ar}\textcolor{black}{ \ \approx\  } 5\times 10^{-12}\ {\rm \textcolor{black}{a}}^{-1}\eqf just at the beginning of that eon. A similar figure for the early Sun's mass loss rate can be inferred from \rfr{ora} and the estimates in \citep{SuzukiEPS2012}. In \citep{lrsp-2004-2} it is argued that the young Sun could not have been more than $0.2\textcolor{black}{\ \%}$ more massive at the beginning of the Archean eon.

On the other hand, in \citep{2000GeoRL..27..501G}  an upper bound of
\eqi\dot M_{\rm \pi^{01}UMa}\textcolor{black}{ \ \approx\   (4 \ {\rm to}\ 5)}\times 10^{-11} M_{\odot}\ {\rm \textcolor{black}{a}}^{-1}\eqf for $\pi^{01}$ Ursa Majoris, a  $0.3$ Ga old solar-mass star, is reported.
Similar figures for other young Sun-type stars have been recently proposed in \citep{2013ApJ...764..170D} as well.

\textcolor{black}{At the post-Newtonian level, general relativity predicts the existence of a test particle acceleration in the case of a time-dependent potential. Indeed, from Eq. (2.2.26) and Eq. (2.2.49) of \citep{1991ercm.book.....B}, written for the case of the usual Newtonian monopole, it can be obtained \citep{2010SRXPh2010.1249I}
\eqi\bds{A}_{\rm GR}  = -3\rp{\dot\mu}{c^2 r}\bds v,\lb{ger}\eqf where $\mu\doteq GM$. The orbital consequences of \rfr{ger} were worked out in \citep{2010SRXPh2010.1249I}:
a secular increase of the distance
\eqi \dot r = -\rp{6\dot\mu}{c^2} + \mathcal{O}\ton{e^2}\lb{mini} \eqf
occurs. It is completely negligible, even for figures as large as \rfr{necess} by assuming that the change in $\mu$ is due to the mass variation.
Indeed, \rfr{mini}, calculated with \rfr{necess}, yields a distance rate as little as $\dot r \textcolor{black}{ \ \approx\  } 3\times 10^{-7}\ $m \textcolor{black}{a}$^{-1}$.
}
\subsection{Non-isotropic mass loss of the Earth due to a possible erosion of its hydrosphere driven by the solar wind}

Let us, now, examine the other potential source of the reduction of the strength of the gravitational interaction in the Sun-Earth system, i.e. the secular mass loss  of the Earth itself, likely due to the erosion of its fluid component steadily hit by the solar wind. To the best of our knowledge, such a possibility has never been treated in the literature so far.

\textcolor{black}{Let us recall that a body acquiring or ejecting mass due to typically non-gravitational interactions with the surrounding environment experiences the following acceleration \citep{Mes897,Somm52,1963Icar....2..440H,Hadji67,1985AZh....62.1175R,1992CeMDA..53..227P}
\eqi \dert{\bds v}{t} =  \rp{\bds F}{m} + \ton{\rp{\dot m}{m}}\bds u \lb{rocket}\eqf with respect to some inertial frame $\mathcal{K}$. In \rfr{rocket}, $\bds F$ is the sum of all the external forces, while
\eqi\bds u \doteq {\bds V}_{\rm esc} -\bds v \lb{fuga}\eqf is the velocity of the escaping mass with respect to the barycenter of the body. In \rfr{fuga}, ${\bds V}_{\rm esc}$ is is the velocity of the escaping particle with respect to the inertial frame $\mathcal{K}$, and $\bds v$ is the velocity of that point of the body which instantaneously coincides with the body's center of mass; it is referred to $\mathcal{K}$, and does not include the geometric shift of the center of mass caused by the mass loss. If the mass loss is isotropic with respect to the body's barycenter, then the second term in \rfr{rocket}
vanishes.}

\textcolor{black}{In the case of a star-planet system \citep{1985AZh....62.1175R}, $\bds F$ is the usual Newtonian gravitational monopole, and the mass loss is anisotropic; moreover, ${\bds V}_{\rm esc}$ is radially directed from the star to the planet.}
According to \citep{2012NewA...17..356I}, the orbital effect  on the distance $r$ is
\eqi\rp{\dot r(t)}{r(t)} = - 2\rp{\dot m}{m},\lb{terramassa}\eqf where it was assumed that the characteristic timescale of the generally time-dependent percent mass loss rate is much larger than the orbital period. \textcolor{black}{It is worthwhile noticing that \rfr{terramassa} does not depend on $V_{\rm esc}$; it is the outcome of a perturbative calculation with the Gauss equations in which no approximations concerning $v$ and $V_{\rm esc}$ were assumed \citep{2012NewA...17..356I}. The eccentricity $e$, the inclination $I$ and the node $\Om$ do not secularly change, while the pericenter $\omega$ undergoes a secular precession depending on $V_{\rm esc}$ \citep{2012NewA...17..356I}.}
If \rfr{rate} was entirely due to \rfr{terramassa}, then the hypothesized Earth mass loss rate would be as large as
\eqi\rp{\dot m}{m}\textcolor{black}{ \ \approx\  } -1.7\times 10^{-11}\ {\rm \textcolor{black}{a}^{-1}},\ 0.77\ {\rm Ga}\leq t\leq 2.07\ {\rm Ga}.\lb{omamma}\eqf
It implies that, at the beginning of the Archean, the Earth was more massive than now by $\textcolor{black}{ \ \approx\  } 2\textcolor{black}{\ \%}$. Thus,  by keeping the solid part of the Earth unchanged, its fluid part should have been larger than now by the non-negligible amount
\eqi \Delta m_{\rm fl} = 0.02 m_0^{\rm tot}.\lb{minchia}\eqf
For a comparison, the current mass  of the fluid part of the Earth is largely dominated by the hydrosphere, which, according to http://nssdc.gsfc.nasa.gov/planetary/factsheet/earthfact.html, amounts to
\eqi m_0^{\rm hy} = 1.4\times 10^{21}\ {\rm kg} = 2.3\times 10^{-4}  m_0^{\rm tot};\lb{atmomass}\eqf
the current mass of the Earth's atmosphere is 274 times smaller than \rfr{atmomass}.
Also for such a postulated mechanism, it should be checked if it is realistic in view of the present-day knowledge. To this aim, it should be recalled that the fluid part of the Earth at the beginning of the Archean eon is the so-called \virg{second atmosphere} \citep{Zahnle010}, and that, to an extent which is currently object of debate \citep{2007Earth...2...43M}, it should have been influenced by  the Terrestrial Late Heavy Bombardment (TLHB) \citep{2000orem.book..475R}   \textcolor{black}{$ \ \approx\   (4\ {\rm to}\ 3.8)$ Ga} ago. In particular, in regard to the composition of the Earth's atmosphere, it is crucial to realistically asses if the extraterrestrial material deposited during the TLHB was mainly constituted of cometary matter or chondritic (i.e. asteroidal) impactors \citep{2013A&A...551A.117B}. Another issue to be considered is if the spatial environment of the Earth could allow for a hydrospheric/atmospheric erosion as large as \rfr{omamma}. To this aim, it is important to remark that the terrestrial magnetic field, which acts as a shield from the eroding solar wind, was only \citep{2010Sci...327.1238T} \textcolor{black}{$ \ \approx\   50\ \%$ to $70\ \%$} of its current level \textcolor{black}{$(3.4\ {\rm to}\ 3.45)$ Ga} ago. Moreover, as previously noted, the stronger stellar wind of the young Sun had consequences on the loss of volatiles and water from the terrestrial early atmosphere \citep{2011JGRA..116.1217S}.
\section{Conclusions}\lb{conclude}

In this paper, we assumed that, given a solar luminosity as little as $75\textcolor{black}{\ \%}$ of its current value at the beginning of the Archean $3.8$ \textcolor{black}{Ga} ago, the Earth was closer to the Sun than now by $4.4\textcolor{black}{\ \%}$ in order to allow for an irradiance large enough to keep a vast liquid ocean on the terrestrial surface. As a consequence, \textcolor{black}{under the assumption that non-climatic effects can solve the Faint Young Sun paradox,}
some physical mechanism should have subsequently move\textcolor{black}{d} our planet to its present-day heliocentric distance in such a way that the solar irradiance stayed substantially constant  during the entire Archean eon, i.e. from $3.8$ \textcolor{black}{Ga} to $2.5$ \textcolor{black}{Ga} ago.

Although it turns out that a relative orbital recession rate of roughly the same order of magnitude of the value of the Hubble parameter $3.8$ \textcolor{black}{Ga} ago would have been required, standard general relativity rules out cosmological explanations for such a hypothesized orbit widening both at the Newtonian and the post-Newtonian level. Indeed,  at the Newtonian level, the first non-vanishing cosmological acceleration is quadratic in the Hubble parameter and, in view of its analytical form, it does not cause any secular variation of the relative distance in a localized two-body system. At the post-Newtonian level, a cosmological acceleration linear in the Hubble parameter has been, in fact, recently predicted. Nonetheless, if, on the one hand, it induces the desired orbital recession, on the other hand, its magnitude, which is determined by well defined ambient parameters such as the speed of light in vacuum, the Hubble parameter and the mass of the primary, is far too small to be of any relevance. Instead, a recently investigated class of modified theories of gravitation violating the strong equivalence principle due to a nonminimal coupling between the matter and the spacetime metric is, in principle, able to explain the putative orbital recession of the Earth. Indeed, it naturally predicts, among other things, also a non-vanishing secular rate of the orbit's semimajor axis  depending on a pair of free parameters whose values can be adjusted to yield just the required one. Moreover, since one of them is, in principle, time-dependent, it can always be assumed that it got smaller in the subsequent 2 \textcolor{black}{Ga} after the end of the Archean in such a way that the current values of the predicted orbit recessions are too small to be detected.

Another physical mechanism of classical origin which, in principle, may lead to the desired orbit expansion is a steady mass loss from either the Sun or the Earth itself. However, such a potentially viable solution presents some difficulties both in terms of the magnitude of the mass loss rate(s) required, especially as far as the Earth's hydrosphere is concerned, and of the timescale itself. Indeed, the Earth should have lost about $2\textcolor{black}{\ \%}$ of its current mass during the Archean.
Moreover, it is generally accepted that a higher mass loss rate for the Sun due to an enhanced solar wind in the past could last for just \textcolor{black}{$(0.2\ {\rm to}\ 0.3)$ Ga} at most.

\textcolor{black}{In conclusion, it is entirely possible  that the Faint Young Sun paradox can be solved by a stronger greenhouse effect on the early Earth; nonetheless, the quest for alternative explanations should definitely be supported and pursued.}


\bibliography{Youngsunbib}{}
\bibliographystyle{mdpi-arXiv}
\end{document}